\documentclass[prl,aps,twocolumn,floats,nofootinbib,showpacs]{revtex4}
\usepackage{amsmath,amssymb,graphicx,psfrag}

\begin{document}
\renewcommand{\ni}{{\noindent}}
\newcommand{\dprime}{{\prime\prime}}
\newcommand{\be}{\begin{equation}}
\newcommand{\ee}{\end{equation}}
\newcommand{\bea}{\begin{eqnarray}} 
\newcommand{\eea}{\end{eqnarray}}
\newcommand{\la}{\langle}
\newcommand{\ra}{\rangle} 
\newcommand{\dg}{\dagger}
\newcommand\lbs{\left[}
\newcommand\rbs{\right]}
\newcommand\lbr{\left(}
\newcommand\rbr{\right)}
\newcommand\f{\frac}
\newcommand\e{\epsilon}
\newcommand\ua{\uparrow}
\newcommand\da{\downarrow}
\title{Many-body mobility edges in a one-dimensional system of interacting fermions}
\author{Sabyasachi Nag and Arti Garg\email{arti.garg@saha.ac.in}} 
\affiliation{Condensed Matter Physics Division, Saha Institute of Nuclear Physics, 1/AF Bidhannagar, Kolkata 700 064, India}
\vspace{0.2cm}
\begin{abstract}
\vspace{0.3cm}
We study many-body localization (MBL) in an interacting one-dimensional system with a deterministic aperiodic  potential. Below the threshold potential $h < h_c$, the non-interacting system has single particle mobility edges (MEs) at $\pm E_c$ while for $ h > h_c$ all the single particle states are localized. We demonstrate that even in the presence of single particle MEs, interactions do not always delocalise the system and the interacting system can have MBL. Our numerical calculation of energy level spacing statistics, participation ratio in the Fock space and Shannon entropy shows that for some regime of particle densities, even for $h < h_c$ many-body states at the top (with $E >E_2$) and the bottom of the spectrum (with $E< E_1$) remain localized though states in the middle of the spectrum are delocalized. Variance of entanglement entropy (EE) also diverges at $E_{1,2}$ indicating a transition from MBL to delocalized regime though transition from volume to area law scaling for EE and from thermal to non-thermal behavior occur inside the MBL regime much below $E_1$ and above $E_2$.
\vspace{0.cm}
\end{abstract} 
\maketitle
Interplay of disorder and interactions in quantum systems is a topic of great interest in condensed matter physics. 
In a non-interacting system with random disorder, any small amount of disorder is sufficient to localize all the single particle states in one and two dimensions~\cite{Anderson,scaling,scaling_rmp}, except in systems where back scattering is suppressed e.g. in graphene~\cite{sdsarma_graphene,agarg}, while in three dimensions (3-d) there occurs a single particle mobility edge (ME) leading to a metal-Anderson Insulator transition. The question of immense interest, that has remained unanswered for decades, is what happens to Anderson localization when both disorder and interactions are present in a system. Recently based on perturbative treatment of interactions for the case where all the single particle states are localised, it has been established that Anderson localization can survive interactions and disordered many-body eigenstates can be localized resulting in a many-body localized (MBL) phase, provided that interactions are sufficiently weak~\cite{Basko}.
The question we want to answer in this work is what happens in the presence of interactions when the non-interacting system has single particle MEs? Conventional wisdom says that in the presence of interactions, localised states will get hybridised with the extended states resulting in delocalization. In this work based on exact diagonalisation (ED) study of an interacting model of spin-less fermions in the presence of a deterministic aperiodic potential, where the non-interacting system has MEs, we demonstrate that for some parameter regimes, many-body states at the top and the bottom of the spectrum remain localised even in the presence of interactions. 

 The MBL phase and the MBL transition are unique for several reasons and challenge the basic foundations of quantum statistical physics~\cite{Huse,Altman}.  In the MBL phase the system explores only an exponentially small fraction of the configuration space and local observables do not thermalize leading to violation of eigenstate thermalisation hypothesis (ETH)~\cite{Deutsch,Srednicki,Rigol}. MBL phase has been shown to have similarity with integrable systems~\cite{Huse2014,shastry} with an extensive number of local integrals of motion~\cite{Abanin,Mueller}. Recently a lot of progress has been made in the field based on numerical analysis of interacting one dimensional models of spin-less fermions or spins with completely random disorder~\cite{Bardarson2014,Bardarson2015,Alet,Sirker,Mueller2016} as well as models where there is no randomness but have a deterministic (quasi-periodic) potential~\cite{Huse2007,Huse2010,Huse2013,Subroto,Sdsarma,Pal1,Pal2,ME1} for example Aubry-Andre (AA) model~\cite{AA} and models with Fibonacci potentials~\cite{Giamarchi} which show a localization to delocalization transition even in 1-d. 

\begin{figure}[h!]
\begin{center}
\vskip-0.35cm
\includegraphics[width=1.75in,angle=-90]{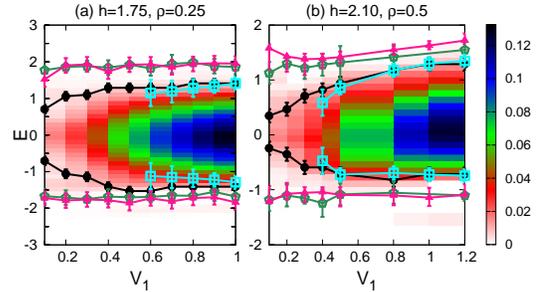}
\vskip-0.4cm
\caption{Phase diagram for model in Eq.\ref{model}. Density plot shows the value of $\eta(E)$ in the thermodynamic limit. Black curves represent MEs $E_{1,2}$ obtained from the level spacing statistics and scaling of NPR and Shannon entropy, while pink, green and sky-blue curves represent transition points from area to volume scaling law of EE, non-thermal to thermal transition points from ETH and points where variance of EE is peaked respectively.} 
\label{pd}
\end{center}
\end{figure}

In this work we consider a 1-d deterministic model which has been explored in context of Anderson localization~\cite{Fishman} and has been shown to have tunable single particle MEs at $\pm E_c$~\cite{Sarma1990} for the strength of aperiodic potential $h < h_c$. Recently this model was studied with interactions ~\cite{Subroto} and it was concluded that the model does not show MBL for $h < h_c$. We demonstrate, based on a careful energy resolved analysis, that for a broad range of parameters, interactions do not delocalize the entire spectrum. Main findings of our work, presented in the phase diagram of Fig.~\ref{pd}, are following. For $h<h_c$ with system less than half-filled, such that in the non interacting system the Fermi energy $E_f$ is sufficiently below $-E_c$, the interacting system has localized many-body states with energy density $E <E_1$ and $E > E_2$ while the intermediate energy states with $E_1<E<E_2$ are delocalized resulting in two MEs. The characteristic energy scales $E_1$ and $E_2$ are obtained from analysis of normalized participation ratio (NPR) in the Fock space, energy level spacing statistics and Shannon entropy. To the best of our knowledge, it is for the first time that Shannon entropy has been used for characterization of MBL phase. These MEs are also consistent with the transition points at which variance of entanglement entropy (EE) diverges in the thermodynamic limit. However, crossover from area to volume law scaling of EE happens at $\tilde{E}_{1,2}$ inside the localised regime with a broad regime of MBL states obeying ETH and the volume law scaling of EE specially for weak interactions.

The model we study has the Hamiltonian of the form 
\bea
H=-t\sum_{i}[c^\dagger_ic_{i+1}+h.c.] + \sum_i h_i n_i + V_1\sum_i n_in_{i+1}
\label{model}
\eea
Here $t$ is the nearest neighbor hopping amplitude for spin-less fermions on a $1-d$ chain, $V_1$ is the nearest neighbor repulsion and $h_i$ is the on site potential of form $h_i=h \cos(2\pi\alpha i^n+\phi)$ where $\alpha$ is an irrational number and $\phi$ is an offset. Note that $n=1$ for $V_1=0$ corresponds to AA model which has all single particle states delocalized for $h < 2t$ but in this work we study this model for $n < 1$ for which the system has single particle MEs at $E_c=\pm|2t-h|$ for $h < 2t$~\cite{Sarma1990}. For $h>2t$, all the single particle states are localized for any value of $n$. The phase diagram shown in Fig.~\ref{pd} has been obtained by solving this model using ED on finite size chains with open boundary conditions.
Below we present analysis of various quantities used to obtain the phase diagram of Fig.~\ref{pd}, mainly for two parameters; $h=1.75t$, which is less than $h_c$, for a quarter-filled system($\rho=0.25$) and $h=2.1t$, which is just above $h_c$, at half-filling ($\rho=0.5$). All the data presented below is for $\alpha=\frac{\sqrt{5}-1}{2}$, $n=0.5$ and is averaged over 100-150 configurations obtained by varying the offset $\phi$. 
\begin{figure}[h!]
\begin{center}
\hskip-0.35cm
\includegraphics[width=2.0in,angle=-90]{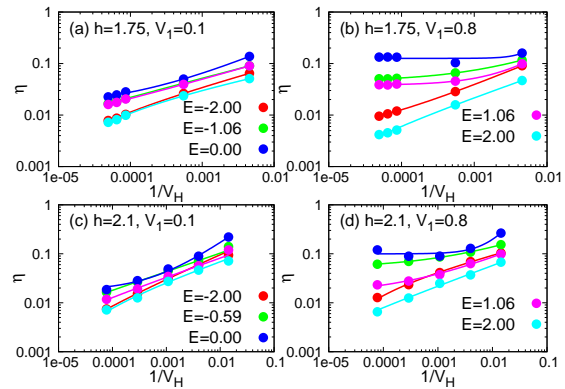}
\vskip-0.3cm
\caption{$\eta(E)$ vs $1/V_H$ for various values of $E$. Here for $V_1=0.1t$, $\eta(E) \sim b*V_H^{-c}$ for all values of $E$ except for a few states with $E \sim 0$ where  $\eta(E)\sim a+b*V_H^{-c}$ .
On increasing $V_1$ more many-body states get delocalized as indicated in panel (b).}
\label{NPR1}
\end{center}
\end{figure}
\\
{\it{Normalized Participation ratio in Fock space(NPR)}}:
We calculate the NPR $\eta(E)$ which represents the fraction of configuration space participating in a many-body state and is defined as
\be
\eta(E) = \frac{1}{\langle \sum_{i,n} |\Psi_n(i)|^4 \delta(E-E_n) \rangle_C V_H}
\ee
where $\Psi_n(i)$ is an eigenfunction with eigenvalue $E_n$ of the Hamiltonian in Eq.~\ref{model}, $V_H$ is the volume of the Fock space, and $\langle \rangle _C$ indicates the configuration averaging. Fig.~\ref{NPR1} shows scaling of $\eta(E)$ w.r.t $1/V_H$ for $h=1.75t$ for almost quarter filled system ($\rho \sim 0.25$). For very low and high energy states, $\eta(E) \sim bV_H^{-c}$ and goes to zero in the thermodynamic limit indicating localized nature of these many-body states. For states in the middle of the spectrum $\eta(E) \sim a+b(1/V_H)^c$ with finite value of $a$ in the thermodynamic limit indicating the ergodic nature of these states. From the extrapolated values of $\eta(E)$ in the thermodynamic limit, shown in the density plot of Fig.~\ref{pd}, we obtained two transitions from MBL states to delocalized states at energy densities $E_1$ and $E_2$ (shown in black curves in Fig.~\ref{pd}) such that states with $E < E_1$ and $E> E_2$ are localized while states with $E_1<E<E_2$ are extended. For $h=2.1t$ and $\rho=0.5$, we get a similar picture from the analysis of NPR, shown in the bottom panel of Fig.~\ref{NPR1} except that here at $V_1=0$ all the many-body states are localised.
\begin{figure}[h!]
\begin{center}
\hskip-0.35cm
\includegraphics[width=1.65in,angle=-90]{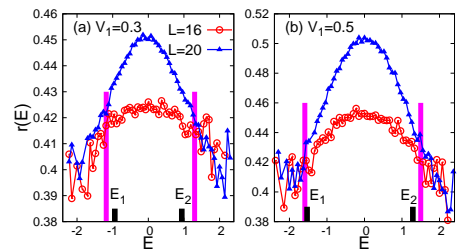}
\vskip-0.6cm
\caption{Ratio of successive gaps $r(E)$ vs $E$ for $h=1.75t$ and $\rho=0.25$ for different system sizes. For $E < E_1$ and $E >E_2$, $r(E)$ is close to its value for PS and does not increase with $L$. But for intermediate energy values $r(E)$ increases with $L$ and approaches the value for WDS.}
\label{r}
\end{center}
\end{figure}

{\it{ Level spacing statistics}}:
The distribution of energy level spacings is expected to follow Poisson statistics (PS) for localized phase while it follows Wigner-Dyson statistics (WDS) for the ergodic phase~\cite{Mehta}. 
We calculate the ratio of successive gaps in energy levels $r_n=\frac{min(\delta_n,\delta_{n+1})}{max(\delta_n,\delta_{n+1})}$~\cite{Huse2007} with $\delta_n=E_{n+1}-E_{n}$ at a given Eigen energy $E_n$ of the Hamiltonian. For PS, the disorder averaged value of $\la r \ra$ is $2ln2-1\approx 0.386$; while for the WDS $\langle r\rangle \approx 0.5295$. 
As shown in Fig.~\ref{r}, for energy states at the bottom and top of the spectrum, $r(E)$ is close to the PS value and does not change significantly with the system size ~\cite{footnote} while in the middle of the spectrum, $r(E)$ increases with the system size approaching the WDS value indicating delocalized nature of these states. Characteristic energy scales obtained are shown in Fig.~\ref{r} which are very close to the MEs $E_{1,2}$ obtained from NPR. 
Plots for $r$, averaged over the entire spectrum~\cite{supplemental} show that for $h \gg h_c$ there exists an  infinite temperature MBL phase where all the many-body states are localised.

\begin{figure}[h!]
\begin{center}
\hskip-0.6cm
\includegraphics[width=2.0in,angle=-90]{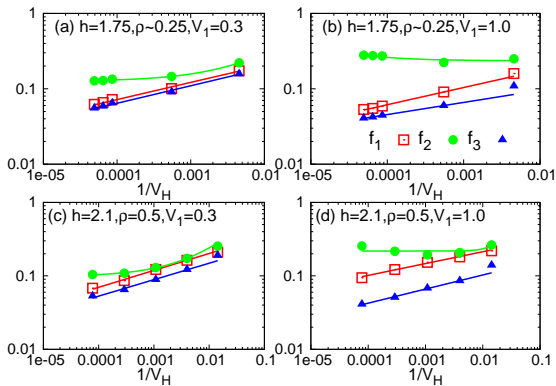}
\vskip-0.3cm
\caption{Scaling of $f_1$, $f_2$ and $f_3$ vs $1/V_H$. Both for $V_1=0.3t$ and $V_1=1.0t$, $f_1$ and $f_3$ vanish in the thermodynamic limit indicating localised nature of states while $f_2$ stays finite indicating the delocalized nature of states in the intermediate energy range.}
\label{shannon1}
\end{center}
\end{figure}

{\it{Shannon Entropy}}:
To further analyze the ergodicity of many-body states, we calculate Shannon entropy for every eigenstate $S(E_n)=-\sum_{i=1}^{V_H}|\Psi_n(i)|^2\ln|\Psi_n(i)|^2$. For a many-body state which gets contribution from all the basis states in the Fock space $S(E_n)\sim \ln(V_H)$ and thus $f(E_n)=\exp(S(E_n))/V_H \sim 1$ while for a localized state which gets significant contribution only from a fraction $N_l$ of the basis states, $f(E_n)\sim N_l/V_H$ and vanishes to zero in the thermodynamic limit.  Fig.~\ref{shannon1} shows the scaling of $f_{1,2,3}$ which are obtained by averaging $f(E_n)$ over three regions of the spectrum, namely, $E_n< E_1$, $E_1<E_n<E_2$ and $E_n>E_2$ obtained from NPR analysis. In the thermodynamic limti, $f_2$ is finite while $f_1$ and $f_3$ vanish~\cite{supplemental} indicating the transition from MBL to delocalized states across the spectrum. 

\begin{figure}[h!]
\begin{center}
\vskip-1cm
\hskip-0.5cm
\includegraphics[width=2.0in,angle=-90]{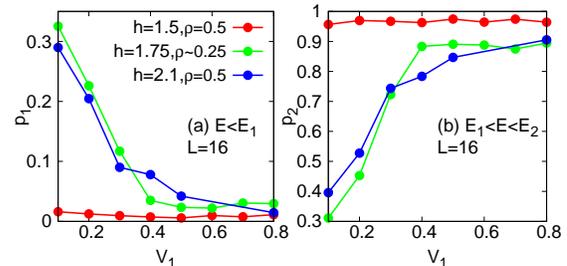}

\vskip-1cm
\caption{Left panel shows $p_{1}$ vs $V_1$ for various values of $h$ and $\rho$. Note that for $h=1.75t,\rho=0.25$, $p_1$ is finite while for $h=1.5t$ and $\rho=0.5$, $p_1$ is vanishingly small. For $h > h_c$, even at half-filling there is a finite fraction of states in the localized sector. Right panel shows $p_2$, fraction of states in the delocalized part of the spectrum. }
\label{frac}
\end{center}
\vskip-0.5cm
\end{figure}
\vskip0.1cm
{\it{Comparison of $h < h_c$ quarter-filled and half-filled case}}:
So far we presented results for $h < h_c$ quarter-filled case.
Now consider the half-filled system ($\rho=0.5$) for $h <h_c$ where for the non-interacting case $-E_c<E_f<E_c$. For $V_1\ne 0$, one again finds $\eta(E)$ and $f(E)$ vanishing in the thermodynamic limit for $E < E_1$ and $E> E_2$ ~\cite{supplemental2}. Further we calculated the fraction of states $p_1 = \frac{N_1}{V_H}$ with $N_1$ being the number of states below $E_1$ and similarly, $p_2$ is the fraction of states in the middle part of spectrum $E_1< E< E_2$ and $p_3=1-p_1-p_2$. As shown in Fig.~\ref{frac} for $h < h_c$, $p_1$ for the system with $\rho=0.5$ is vanishingly small and is much smaller than its value for the system with $\rho=0.25$. This indicates that there is no MBL phase for $h <h_c$ half-filled system which is consistent with earlier results~\cite{Subroto} while for $\rho=0.25$ case presented above, a finite fraction of many body spectrum is localized. 

\begin{figure}[h!]
\begin{center}
\hskip-0.5cm
\includegraphics[width=1.5in,angle=-90]{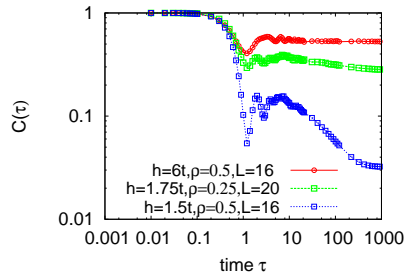}
\vskip-0.2cm
\caption{Return probability $C(\tau)$ for $V_1=1.0t$. In the large time limit, $C(\tau)$ stays finite and constant for $h=6t$ indicating localised nature of the system while for $h=1.5t$, $C(\tau)$ decreases with $\tau$ indicating its delocalised nature. For $h=1.75t$ and $\rho=0.25$, $C(\tau)$ decreases very slowly indicating localised nature of most of the spectrum.}
\label{ct}
\end{center}
\vskip-0.5cm
\end{figure}

{\it{Return Probability}}:
Next we calculate the probability of return of particles to their initial position $i$ as a function of time $C_i(\tau)=\f{1}{V_H}\sum_m\la \Psi_m|(n_i(\tau)-\f{1}{2})(n_i(0)-\f{1}{2})|\Psi_m\ra$, which is averaged over all the sites to obtain $C(\tau)$. For $h=6t \gg h_c$, where all the many body states are localised, $C(\tau)$ remains constant with time in the large time limit as shown in Fig.~\ref{ct}. For $h<h_c$ and $\rho=0.5$, where the system remains delocalised even in the presence of interactions, $C(\tau)$ decays with time in the large time limit. On the contrary for $h <h_c$ quarter-filled case where there are two many-body MEs, $C(\tau)$ decreases very slowly (remaining almost constant with time) and has value much larger than that for $h < h_c$ half-filled system which indicates that interactions have not delocalised the entire many-body spectrum. 

\begin{figure}[h!]
\begin{center}
\includegraphics[width=1.8in,angle=-90]{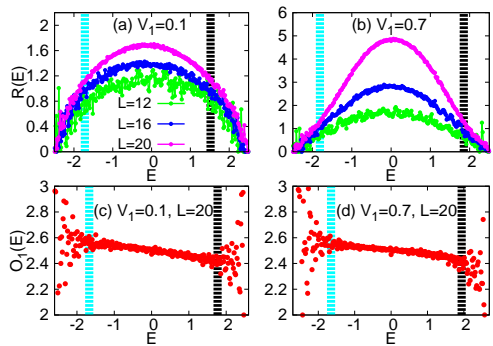}
\caption{Top panel shows $R(E)$ vs $E$. For $E< \tilde{E}_1$ and $E > \tilde{E}_2$, $R(E)$ is same for all $L$ but for the intermediate states $R(E)$ increases with $L$ indicating their ergodic nature. 
The bottom panel shows $O(E)$ vs $E$ which shows large fluctuations in its value for near by eigenstates for $E<\tilde{\tilde{E}}_1\sim \tilde{E}_1$ and $E>\tilde{\tilde{E}}_2\sim \tilde{E}_2$. This data is for $h=1.75t$ with $\rho=0.25$. Corresponding figure for $h=2.1t$ is shown in supplement~\cite{supplemental}.}
\label{Renyi1}
\vskip-1.8cm
\end{center}
\end{figure}

{\it{Entanglement Entropy and ETH}}:
Entanglement entropy (EE) is a useful tool to distinguish between the ergodic and many-body localized phases. We divide the lattice into two subsystems A and B of sites $L/2$ and calculate the Renyi entropy $R(E_n) = -log[Tr_A \rho_A(E_n)^2]$ where $\rho_A$ is the reduced density matrix obtained by integrating the total density matrix $\rho_{total}(E_n) = |\Psi_n\ra \la \Psi_n|$ over the degree of freedom of subsystem B.  EE is expected to obey the volume law of scaling $R\sim L^d$ in the ergodic phase while it is suppressed for the MBL phase showing an area law scaling $R\sim L^{d-1}$~\cite{Nayak,Huse2013,Sdsarma} with $d=1$ for the model under study. 
As shown in Fig.~\ref{Renyi1}, for low and high energy states average $R(E)$ is same for various $L$ values indicating that the states in this energy regime are localized while for states in the middle of the spectrum $R(E)$ increases with $L$ indicating their delocalised nature though the transition points $\tilde{E}_{1,2}$ obtained from EE are little off from $E_{1,2}$ as shown in Fig.~\ref{pd}. This indicates that there is a regime where many-body states are localized in the Fock space but still EE increases with $L$. Though NPR overestimates the MBL regime specially for weak interactions~\cite{Sdsarma} having zero thermodynamic value for partially extended states, EE might give an over estimate of extended regime. The EE data for non-interacting case shows ~\cite{supplemental2} that even for many-body states, which are slater determinants of all localised single particle states and  hence are definitely localised, $R$ increases with the system size. Hence the actual many-body ME lies somewhat little below $E_1$ but definitely above $\tilde{E}_1$. 

To check for the ETH in various parameter regimes we calculated expectation value of the number operator on subsystem A which has $L/2$ sites w.r.t various eigenstates. We define $\hat{O} = \sum_{i=1}^{L/2} \hat{n}_i$ where $\hat{n}_i$ is the number operator for spin-less fermions at site $i$. As shown in bottom panel of Fig.~\ref{Renyi1}, for $E < \tilde{\tilde{E}}_1$ and $E > \tilde{\tilde{E}}_2$, many-body system is not thermal showing large fluctuations in $O(E)$ for nearby energy states  while for $\tilde{\tilde{E}}_1 < E< \tilde{\tilde{E}}_2$, system is ergodic and obeys ETH. As shown in Fig.~\ref{Renyi1}, $\tilde{\tilde{E}}_{1,2} \sim \tilde{E}_{1,2}$ within numerical errors.

\begin{figure}[h!]
\begin{center}
\vskip-0.6cm
\includegraphics[width=1.8in,angle=-90]{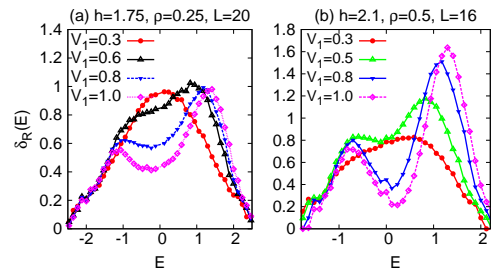}
\vskip-0.7cm
\caption{Variance $\delta_R(E)$ of EE as a function of $E$ for various values of $V_1$. For $V_1 > 0.3t$, $\delta_R(E)$ shows two clear peaks, indicating localisation to delocalization transition in the many-body spectrum.}
\label{var}
\end{center}
\vskip-0.5cm
\end{figure}

We also calculated the variance of EE $\delta_R(E)=\la R(E)^2\ra-\la R(E) \ra^2$ shown in Fig.~\ref{var}. In the thermodynamic limit $\delta_R$ should be zero deep inside the localized and delocalised phases but at the transition point it diverges due to contribution from both the extended and the localized states~\cite{Bardarson2014} which is reflected as a peak in finite size calculations. Our data shows two clear peaks in $\delta_R(E)$ vs $E$ curve for intermediate values of $V_1$ indicating two transition points which are very close to $E_{1,2}$ as shown in Fig.~\ref{pd} where sky-blue curves represent peak positions in $\delta_R(E)$ vs $E$ curve. Note that for small values of $V_1$ the two peaks in $\delta_R(E)$ vs $E$ curve are very close to each other and it is difficult to identify the peak positions precisely. 

In summary we have analysed MBL in an interacting 1-d model of spin-less fermions in the presence of an aperiodic potential where the non-interacting system has mobility edges at $\pm E_c$ for $h <h_c$. We demonstrated that for the system less than half-filled such that $E_f<-E_c$, the interacting system has two mobility edges with localised states living on the low and very high energy part of the spectrum while the middle of the spectrum has delocalized states. To the best of our knowledge, thermal hot spots, predicted to be the cause for the lack of many-body mobility edges~\cite{Mueller_hs} are not possible in the model we have studied. The question of fundamental interest is why interactions in this model can not delocalise all the many-body states and will be addressed in future work.

{\bf Acknowledgements:}
We would like to acknowledge Subroto Mukerjee for many useful discussions and for clarifications on the issue of delocalization protection.


\begin{thebibliography}{99}
\bibitem{footnote}
 Within the system sizes studied, $r(E)$ vs $E$ data for various $L$ values is a bit noisy on the localised side.
But one can see clear difference in $L$ dependence of $r(E)$ for energy states in the middle of the spectrum where $r$ increases with $L$ approaching WDS value.
The estimate of critical energies from this system size dependence of $r(E)$ is very close to those obtained from couple of other diagnostics.
\end{thebibliography}
\end{document}